\documentclass[10pt]{article}
\usepackage{graphicx}
\usepackage{amsmath}
\usepackage{amssymb}
\usepackage{caption2}
\begin{document}
\begin{center}
{\large {\bf \sc{  Analysis of the $\frac{3}{2}^{\pm}$ pentaquark states in the diquark model with QCD sum rules  }}} \\[2mm]
Jun-Xia Zhang$^{1,2}$, Zhi-Gang  Wang$^{1}$\footnote{Email: zgwang@aliyun.com.}, Zun-Yan Di$^{1,2}$     \\
$^{1}$Department of Physics, North China Electric Power University, Baoding 071003, P. R. China \\
$^{2}$School of Nuclear Sicence and Engineering, North China Electric Power University, Beijing 102206, P. R. China \\
\end{center}

\begin{abstract}

In this article, we construct the scalar-diquark-axialvector-diquark-antiquark type interpolating  currents, and study the masses and pole residues of the
$J^{P}=\frac{3}{2}^{\pm}$ hidden-charmed pentaquark states with the QCD sum rules. In calculations, we use the formula $\mu=\sqrt{M^2_{P}-(2{\mathbb{M}}_c)^2}$ to determine the energy scales of the QCD spectral densities. We obtain the masses of the hidden-charm pentaquark states with the strangeness $S=-1$ and
$S=-2$, which can be confronted to the experimental data in the future.
\end{abstract}



\section{Introduction}

 In 2015, the LHCb collaboration studied the $\Lambda_{b}^{0}\rightarrow J/\psi p$ decays, and observed two exotic hidden-charm pentaquark resonances, $P_c(4380)$ and
$P_c(4450)$, in the  $J/\psi p$  mass spectrum with the significance of more than 9$\sigma$ \cite{LHCb-4380}. They are  good
candidates of pentaquark states, which are made of four quarks and one antiquark. The measured masses and widths are
\begin{eqnarray}
&&M_{P_c(4380)}=(4380\pm 8 \pm 29)\,  { \rm{MeV} },   \,\,\,   \Gamma_{P_c(4380)} = (205\pm18\pm 86) \, \rm{MeV}\, , \\
&&M_{P_c(4450)}=(4449.8\pm 1.7\pm 2.5)\, { \rm{MeV} }, \,\,\, \Gamma_{P_c(4450)}= (39\pm 5\pm 19) \, \rm{MeV}\, .
 \end{eqnarray}
The preferred spin-parity assignments of the $P_c(4380)$ and $P_c(4450)$ are $J^{P}=\frac{3}{2}^{-}$ and $\frac{5}{2}^{+}$,
respectively \cite{LHCb-4380}. The $P_c(4380)$ and $P_c(4450)$ have attracted much attentions of the theoretical physicists, several possible
assignments were suggested, such as the $\Sigma_{c}\bar{D}^{*}$, $\Sigma_{c}^{*}\bar{D}^{*}$, $J/\psi N$(1440), $J/\psi N$(1520)
molecule-like pentaquark states \cite{mole-penta} (or not the
molecular pentaquark states \cite{mole-penta-No}), the diquark-diquark-antiquark type pentaquark states \cite{Maiani1507,WangPc-4380,WangHuang-Pc,WangPc-12,WangNPB-2016,di-di-anti-penta},
the diquark-triquark type pentaquark states \cite{di-tri-penta}, the re-scattering effects \cite{rescattering-penta}, etc.

The QCD sum rules have been applied extensively to study the
hidden-charm (bottom) tetraquark or molecular states \cite{Tetraquark-Mole,WangTetraquark,WangMolecule,WangEPJC4260} and pentaquark
states \cite{WangPc-4380,WangHuang-Pc,WangPc-12,WangNPB-2016}.   We constructed the diquark-diquark-antiquark type interpolating currents to study the $P_c(4380)$ and $P_c(4450)$ with QCD sum rules by calculating the contributions of the vacuum condensates up to dimension-10 in the operator product expansion and using the energy scale formula to determine the ideal energy scales of the QCD spectral densities  \cite{WangPc-4380}, then we  study other hidden-charm pentaquark states with $J^{P}=\frac{1}{2}^{\pm}$, $\frac{3}{2}^{\pm}$ in Refs.\cite{WangHuang-Pc,WangPc-12,WangNPB-2016}.
In summary, we have studied the $S_L-A_H-\bar{c}$ type hidden-charm pentaquark states with $J^{P}=\frac{3}{2}^{-}$, $\frac{5}{2}^{+}$ and strangeness $S=0$ \cite{WangPc-4380},  the $S_L-S_H-\bar{c}$ type, $S_L-A_H-\bar{c}$ type hidden-charm pentaquark states with $J^{P}=\frac{1}{2}^{\pm}$ and strangeness $S=0$ \cite{WangHuang-Pc},
the $A_L-A_H-\bar{c}$ type, $A_L-S_H-\bar{c}$ type hidden-charm pentaquark states with $J^{P}=\frac{1}{2}^{\pm}$ and strangeness $S=0,\,-1,\,-2,\,-3$ \cite{WangPc-12},
the $A_L-A_H-\bar{c}$ type, $A_L-S_H-\bar{c}$ type hidden-charm pentaquark states with $J^{P}=\frac{3}{2}^{\pm}$ and strangeness $S=0,\,-1,\,-2,\,-3$ \cite{WangNPB-2016}, where the $S_{L/H}$ denote the light and heavy scalar diquark states, the  $A_{L/H}$ denote the light and heavy axialvector diquark states.

 In this article, we extend our previous work to study the masses and  pole residues of the  $S_L-A_H-\bar{c}$ type hidden-charm pentaquark states with $J^{P}=\frac{3}{2}^{-}$  and strangeness $S=-1,\,-2$.

The article is arranged as follows: we derive the QCD sum rules for the masses and pole residues of the $J^{P}=\frac{3}{2}^{\pm}$ hidden-charm pentaquark states with strangeness $S=-1$, $-2$ in the Sect.2; in the Sect.3, we present the numerical results and discussions; and Sect.4 is
reserved for our conclusion.

\section{QCD sum rules for the $\frac{3}{2}^{\pm}$ hidden-charm pentaquark states}

Now  we  write down the two-point correlation functions $\Pi^i_{\mu\nu}(p) $  with $i=1,2$,
\begin{eqnarray}
\Pi^i_{\mu\nu}(p)=i\int d^{4} x e^{ip\cdot x}\left\langle 0|T\left\{J^i_{\mu}(x)\bar{ J}^i_{\nu }(0)\right\}|0\right\rangle \, ,
\end{eqnarray}
where
 \begin{eqnarray}
   J_{\mu}^1(x)&=&\varepsilon^{i l a}\varepsilon^{i j k}\varepsilon^{l m n} u_{j}^T (x)C\gamma_{5} s_{k}(x) u_{m}^T (x)C\gamma_{\mu} c_{n}(x)C \bar{c}_{a}^T(x) \, , \\
 J_{\mu}^2(x)&=&\varepsilon^{i l a}\varepsilon^{i j k}\varepsilon^{l m n} u_{j}^T (x)C\gamma_{5} s_{k}(x) s_{m}^T (x)C\gamma_{\mu}c_{n}(x)C \bar{c}_{a}^T(x) \, ,
\end{eqnarray}
the $i$, $j$, $k$, $l$, $m$, $n$ and $a$ are color indices, the $C$ is the charge
conjugation matrix.

At the hadron side, we insert a complete set of intermediate hadronic states with the same quantum numbers as the
current operators $J_{\mu}^{i}(x)$ into the correlation functions  $\Pi^i_{\mu\nu}(p) $ to obtain the hadronic
representation \cite{SVZ79,PRT85}. After isolating the pole terms of the lowest states of the hidden-charm pentaquark states with spin $J=\frac{3}{2}$, we
get the following result,
\begin{eqnarray}
 \notag\
  \Pi^i_{\mu\nu}(p) & = & \lambda^{-2}_{i}  {\!\not\!{p}+ M^i_{-} \over M^{i2}_{-}-p^{2}  } \left(- g_{\mu\nu}+\frac{\gamma_\mu\gamma_\nu}{3}+\frac{2p_\mu p_\nu}{3p^2}-\frac{p_\mu\gamma_\nu-p_\nu \gamma_\mu}{3\sqrt{p^2}}\right)\nonumber\\
&&+  \lambda^{+2}_{i}  {\!\not\!{p}- M^i_{+} \over M_{+}^{i2}-p^{2}  } \left(-
g_{\mu\nu}+\frac{\gamma_\mu\gamma_\nu}{3}+\frac{2p_\mu p_\nu}{3p^2}-\frac{p_\mu\gamma_\nu-p_\nu \gamma_\mu}{3\sqrt{p^2}}
\right)   +\cdots
\nonumber\\
 & =& \Pi^i(p^{2})(-g_{\mu\nu})+\cdots \, ,
\end{eqnarray}
where the $M^i_{\pm}$ are the masses of the lowest pentaquark states with the parity $\pm$, respectively, and the
$\lambda^\pm_i$ are the corresponding pole residues.

We obtain the hadronic spectral densities through  dispersion relation \cite{WangPc-4380} as
\begin{eqnarray}
\frac{{\rm{Im}\Pi}^i(s)}{\pi}&=& p\!\!\!/ \left[{\lambda^{-2}_i}\delta\left(s-M^{i2}_{-}\right)+\lambda^{+2}_i\delta\left(s-M^{i2}_{+}\right)\right]
+\left[M^i_{-}\lambda^{-2}_i\delta\left(s-M^{i2}_{-}\right)-M^{i}_{+}\lambda^{+2}_i\delta\left(s-M^{i2}_{+}\right)\right] \nonumber\\
&=&p\!\!\!/\rho_{H}^{i1}(s)+\rho_{H}^{i0}(s)\, .
\end{eqnarray}
Then we introduce the weight function exp$(-\frac{s}{T^{2}})$ to
obtain the QCD sum rules at the hadron side,
\begin{eqnarray}
&&\int_{4m_{c}^{2}}^{s_{0}}ds
\left[\sqrt{s}\rho_{H}^{i1}(s)+\rho_{H}^{i0}(s)\right]\exp\left(-\frac{s}{T^2}\right)=2M^i_{-}\lambda^{-2}_i\exp\left(-\frac{M^{i2}_{-}}{T^2}\right) \, , \\
&&\int_{4m_{c}^{2}}^{s_{0}}ds\left[\sqrt{s}\rho_{H}^{i1}(s)-\rho_{H}^{i0}(s)\right]\exp\left(-\frac{s}{T^2}\right)=2M^i_{+}\lambda^{+2}_i\exp\left(-\frac{M^{i2}_+}{T^2}\right)\, ,
\end{eqnarray}
where $s_{0}$ are continuum threshold parameters and the  $T^{2}$ are the Borel parameters. The contributions of  the negative
parity hidden-charm pentaquark states are separated from that of the positive
parity.

In the following, we carry out the operator product expansion for
the correlation functions $\Pi^i_{\mu\nu}(p) $ in perturbative
QCD. Contracting the $u$, $s$ and $c$ quark fields in the correlation
functions with Wick theorem, we obtain
\begin{eqnarray}
\Pi_{\mu\nu}^{1}(p)&=&i\,\varepsilon^{ila}\varepsilon^{ijk}\varepsilon^{lmn}\varepsilon^{i^{\prime}l^{\prime}a^{\prime}}\varepsilon^{i^{\prime}j^{\prime}k^{\prime}}
\varepsilon^{l^{\prime}m^{\prime}n^{\prime}}\int d^4x e^{ip\cdot x}CC_{a^{\prime}a}^T(-x)C \nonumber\\
&&\left\{    {\rm Tr}\left[\gamma_5 S_{kk^\prime}(x) \gamma_5 C U^{T}_{jj^\prime}(x)C\right] \, {\rm Tr}\left[\gamma_\mu C_{nn^\prime}(x) \gamma_\nu C U^{T}_{mm^\prime}(x)C\right] \right. \nonumber\\
&&\left.-   {\rm Tr} \left[\gamma_5 S_{kk^\prime}(x) \gamma_5 C
U^{T}_{mj^\prime}(x)C \gamma_\mu C_{nn^\prime}(x) \gamma_\nu C
U^{T}_{jm^\prime}(x)C\right]    \right\} \, ,
\end{eqnarray}
\begin{eqnarray}
\Pi_{\mu\nu}^{2}(p)&=&i\,\varepsilon^{ila}\varepsilon^{ijk}\varepsilon^{lmn}\varepsilon^{i^{\prime}l^{\prime}a^{\prime}}\varepsilon^{i^{\prime}j^{\prime}k^{\prime}}
\varepsilon^{l^{\prime}m^{\prime}n^{\prime}}\int d^4x e^{ip\cdot x}CC_{a^{\prime}a}^T(-x)C \nonumber\\
&&\left\{    {\rm Tr} \left[\gamma_5 S_{km^\prime}(x) \gamma_5 C U^{T}_{jj^\prime}(x)C\right] \, {\rm Tr}\left[\gamma_\mu C_{nn^\prime}(x) \gamma_\nu C S^{T}_{mm^\prime}(x)C\right] \right. \nonumber\\
&&\left.-   {\rm Tr} \left[ C S^{T}_{km^\prime}(x) C \gamma_5
U^{T}_{jj^\prime}(x)\gamma_5 C S^{T}_{mk^\prime}(x) C \gamma_\mu
C_{nn^\prime}(x) \gamma_\nu \right] \right\} \, ,\\
 \notag\
\end{eqnarray}
where the $U_{ij}(x)$, $S_{ij}(x)$ and $C_{ij}(x)$ are the full $u$, $s$
and $c$ quark propagators, respectively,

\begin{eqnarray}
U_{ij}(x)&=&\frac{i\delta_{ij}\!\not\!{x}}{
2\pi^2x^4}-\frac{\delta_{ij}\langle \bar{q}q\rangle}{12}
-\frac{\delta_{ij}x^2\langle \bar{q}g_s\sigma Gq\rangle}{192}
-\frac{ig_sG^{a}_{\alpha\beta}t^a_{ij}(\!\not\!{x}
\sigma^{\alpha\beta}+\sigma^{\alpha\beta}
\!\not\!{x})}{32\pi^2x^2} -\frac{1}{8}\langle\bar{q}_j\sigma^{\mu\nu}q_i \rangle \sigma_{\mu\nu}+\cdots \, ,\nonumber \\
S_{ij}(x)&=&\frac{i\delta _{ij}x \!\!\!/}{2\pi ^{2}x^{4}}-\frac{\delta _{ij}m_{s}}{%
4\pi ^{2}x^{2}}-\frac{\delta _{ij}\left\langle \overline{s}s\right\rangle }{%
12}+\frac{i\delta _{ij}x\!\!\!/ m_{s}\left\langle \overline{s}s\right\rangle }{48}-%
\frac{\delta _{ij}x^{2}\left\langle \overline{s}g_{s}\sigma
Gs\right\rangle }{192}+\frac{i\delta _{ij}x^{2}x\!\!\!/
m_{s}\left\langle \overline{s}g_{s}\sigma Gs\right\rangle
}{1152}  \nonumber\\
&&-\frac{ig_{s}G_{\alpha \beta }^{a}t_{ij}^{a}(x\!\!\!/ \sigma
^{\alpha \beta }+\sigma ^{\alpha \beta }x\!\!\!/)}{32\pi
^{2}x^{2}}-\frac{\left\langle \overline{s}_{j}\sigma ^{\mu \nu
}s_{i}\right\rangle \sigma _{\mu \nu }}{8}+\cdots \, ,
\end{eqnarray}

\begin{eqnarray}
C_{ij}(x)&=&\frac{i}{(2\pi)^4}\int d^4k e^{-ik \cdot x} \left\{
\frac{\delta_{ij}}{\!\not\!{k}-m_c}
-\frac{g_sG^n_{\alpha\beta}t^n_{ij}}{4}\frac{\sigma^{\alpha\beta}(\!\not\!{k}+m_c)+(\!\not\!{k}+m_c)
\sigma^{\alpha\beta}}{(k^2-m_c^2)^2}\right.\nonumber\\
&&\left. -\frac{g_s^2 (t^at^b)_{ij} G^a_{\alpha\beta}G^b_{\mu\nu}(f^{\alpha\beta\mu\nu}+f^{\alpha\mu\beta\nu}+f^{\alpha\mu\nu\beta}) }{4(k^2-m_c^2)^5}+\cdots\right\} \, ,\nonumber\\
f^{\alpha\beta\mu\nu}&=&(\!\not\!{k}+m_c)\gamma^\alpha(\!\not\!{k}+m_c)\gamma^\beta(\!\not\!{k}+m_c)\gamma^\mu(\!\not\!{k}+m_c)\gamma^\nu(\!\not\!{k}+m_c)\,
,
\end{eqnarray}
where $t^{n}=\frac{\lambda^{n}}{2}$, the $\lambda^{n}$ is the
Gell-Mann matrix \cite{PRT85}. Then we compute the integrals both
in the coordinate and momentum spaces, and obtain the correlation
functions $\Pi^{i}_{\mu\nu}(p) $ , therefore the QCD spectral
densities $\rho_{QCD}^{i1}(s)$ and $\tilde{{\rho}}_{QCD}^{i0}(s)$
at the quark level through dispersion relation,

Finally,  we take the quark-hadron duality below the continuum thresholds $s_{0} $ and introduce the weight function
$\exp(-\frac{s}{T^{2}})$ to obtain the following QCD sum rules:
\begin{eqnarray}
&&2M^i_{-}\lambda^{-2}_i\exp\left(-\frac{M^{i2}_{-}}{T^{2}}\right)=\int_{4m_{c}^{2}}^{s_{0}}ds \left[\sqrt{s}\rho_{QCD}^{i1}(s)+m_c\tilde\rho_{QCD}^{i0}(s)\right]\exp\left(-\frac{s}{T^{2}}\right),
\end{eqnarray}
\begin{eqnarray}
&&2M^i_{+}\lambda^{+2}_i\exp\left(-\frac{M^{i2}_{+}}{T^{2}}\right)=\int_{4m_{c}^{2}}^{s_{0}}ds
\left[\sqrt{s}\rho_{QCD}^{i1}(s)-m_c\tilde\rho_{QCD}^{i0}(s)\right]\exp\left(-\frac{s}{T^{2}}\right),
\end{eqnarray}
where
\begin{eqnarray}
\rho_{QCD}^{i1}(s)=&\rho_{0}^{i1}(s)+\rho_{3}^{i1}(s)+\rho_{4}^{i1}(s)+\rho_{5}^{i1}(s)+\rho_{6}^{i1}(s)+\rho_{8}^{i1}(s)+\rho_{9}^{i1}(s)+\rho_{10}^{i1}(s)\, ,\nonumber\\
\tilde{{\rho}}_{QCD}^{i0}(s)=&\tilde{{\rho}}_{0}^{i0}(s)+\tilde{{\rho}}_{3}^{i0}(s)+\tilde{{\rho}}_{4}^{i0}(s)+\tilde{{\rho}}_{5}^{i0}(s)+\tilde{{\rho}}_{6}^{i0}(s)+\tilde{{\rho}}_{8}^{i0}(s)+\tilde{{\rho}}_{9}^{i0}(s)+\tilde{{\rho}}_{10}^{i0}(s)\,
,
\end{eqnarray}
the explicit expressions of the QCD spectral densities $\rho_{j}^{i1}(s)$ and $\tilde{{\rho}}_{j}^{i0}(s)$
with $j=0, 3,4,5,6,8,9,10$ are given  in the appendix.

We differentiate  Eqs.(15-16) with respect to $\frac{1}{T^{2}}$, then eliminate the pole residues $\lambda^{\mp}_i$,  and obtain the QCD sum rules for the
masses of the hidden-charm  pentaquark states.
\begin{eqnarray}
M^{i2}_{-}&=&-\frac{\frac{d}{d(1/T^2)}\int_{4m_{c}^2}^{s_0}ds
\left[\sqrt{s}\rho_{QCD}^{i1}(s)+m_{c}\tilde{\rho}_{QCD}^{i0}(s)\right]\exp\left(-\frac{s}{T^2}\right)}{\int_{4m_{c}^{2}}^{s_{0}}ds
\left[\sqrt{s}\rho_{QCD}^{i1}(s)+m_{c}\tilde{\rho}_{QCD}^{i0}(s)\right]\exp\left(-\frac{s}{T^2}\right)}\, , \\
M^{i2}_{+}&=&-\frac{\frac{d}{d(1/T^2)}\int_{4m_{c}^2}^{s_0}ds
\left[\sqrt{s}\rho_{QCD}^{i1}(s)-m_{c}\tilde{\rho}_{QCD}^{i0}(s)\right]\exp\left(-\frac{s}{T^2}\right)}{\int_{4m_{c}^{2}}^{s_{0}}ds
\left[\sqrt{s}\rho_{QCD}^{i1}(s)-m_{c}\tilde{\rho}_{QCD}^{i0}(s)\right]\exp\left(-\frac{s}{T^2}\right)}\, .
\end{eqnarray}
Once the masses $M^i_{-}$ $(M^i_{+})$ are obtained, we can take them as input
parameters and obtain the pole residues $\lambda^{-}_i$ $(\lambda^{+}_i)$ from the
QCD sum rules in Eqs.(14) and (15).

\section{Numerical results and discussions}
The input parameters  are taken to be the standard values
$\langle\bar{q}q \rangle=-(0.24\pm 0.01\, \rm{GeV})^3$,  $\langle\bar{s}s \rangle=(0.8\pm0.1)\langle\bar{q}q \rangle$,
 $\langle\bar{q}g_s\sigma G q \rangle=m_0^2\langle \bar{q}q \rangle$,  $\langle\bar{s}g_s\sigma G s \rangle=m_0^2\langle \bar{s}s \rangle$,
$m_0^2=(0.8 \pm 0.1)\,\rm{GeV}^2$, $\langle \frac{\alpha_s
GG}{\pi}\rangle=(0.33\,\rm{GeV})^4 $    at the energy scale  $\mu=1\, \rm{GeV}$
\cite{SVZ79,PRT85,ColangeloReview}, $m_{c}(m_c)=(1.275\pm0.025)\,\rm{GeV}$ and $m_s(\mu=2\,\rm{GeV})=(0.095\pm0.005)\,\rm{GeV}$
 from the Particle Data Group \cite{PDG}. Furthermore, we set $m_u=m_d=0$ due to the small current quark masses.
 We take into account
the energy-scale dependence of  the input parameters from the renormalization group equation,
\begin{eqnarray}
\langle\bar{q}q \rangle(\mu)&=&\langle\bar{q}q \rangle(Q)\left[\frac{\alpha_{s}(Q)}{\alpha_{s}(\mu)}\right]^{\frac{4}{9}}\, ,\nonumber\\
\langle\bar{s}s \rangle(\mu)&=&\langle\bar{s}s \rangle(Q)\left[\frac{\alpha_{s}(Q)}{\alpha_{s}(\mu)}\right]^{\frac{4}{9}}\, , \nonumber\\
\langle\bar{q}g_s \sigma Gq \rangle(\mu)&=&\langle\bar{q}g_s \sigma Gq \rangle(Q)\left[\frac{\alpha_{s}(Q)}{\alpha_{s}(\mu)}\right]^{\frac{2}{27}}\, , \nonumber\\
\langle\bar{s}g_s \sigma Gs \rangle(\mu)&=&\langle\bar{s}g_s \sigma Gs \rangle(Q)\left[\frac{\alpha_{s}(Q)}{\alpha_{s}(\mu)}\right]^{\frac{2}{27}}\, , \nonumber\\
m_c(\mu)&=&m_c(m_c)\left[\frac{\alpha_{s}(\mu)}{\alpha_{s}(m_c)}\right]^{\frac{12}{25}} \, ,\nonumber\\
m_s(\mu)&=&m_s({\rm 2GeV} )\left[\frac{\alpha_{s}(\mu)}{\alpha_{s}({\rm 2GeV})}\right]^{\frac{4}{9}} \, ,\nonumber\\
\alpha_s(\mu)&=&\frac{1}{b_0t}\left[1-\frac{b_1}{b_0^2}\frac{\log t}{t} +\frac{b_1^2(\log^2{t}-\log{t}-1)+b_0b_2}{b_0^4t^2}\right]\, ,
\end{eqnarray}
  where $t=\log \frac{\mu^2}{\Lambda^2}$, $b_0=\frac{33-2n_f}{12\pi}$, $b_1=\frac{153-19n_f}{24\pi^2}$, $b_2=\frac{2857-\frac{5033}{9}n_f+\frac{325}{27}n_f^2}{128\pi^3}$,  $\Lambda=213\,\rm{MeV}$, $296\,\rm{MeV}$  and  $339\,\rm{MeV}$ for the flavors  $n_f=5$, $4$ and $3$, respectively  \cite{PDG}, and evolve all the input parameters to the optimal energy scales  $\mu$ to extract the masses of the hidden-charm pentaquark states.

In previous works, we  studied the energy scale dependence of the QCD sum rules for the  hidden-charm (hidden-bottom) tetraquark states and molecular states $X$, $Y$, $Z$  in details for the first time, and suggested an energy scale formula $\mu=\sqrt{M^2_{X/Y/Z}-(2{\mathbb{M}}_Q)^2}$ with the effective heavy quark masses ${\mathbb{M}}_Q$ to determine the ideal energy scales of the QCD spectral densities  \cite{WangTetraquark,WangMolecule}, then we  extended the energy scale formula to study the hidden-charm pentaquark states \cite{WangPc-4380,WangHuang-Pc,WangPc-12,WangNPB-2016} and obtained satisfactory results. In this article,  we use the energy scale formula $\mu=\sqrt{M^2_{P}-(2{\mathbb{M}}_c)^2}$ to determine the energy scales of the QCD spectral densities, and take the updated value of the effective $c$-quark mass ${\mathbb{M}}_c=1.82\,\rm{GeV}$ \cite{WangEPJC4260}. For detailed discussions about the energy scale formula, one can consult Ref.\cite{Wang-Xi3080}.

In Refs.\cite{WangPc-4380,WangHuang-Pc,WangPc-12,WangNPB-2016}, we take the continuum threshold parameters as $\sqrt{s_{0}}=M_{P}+(0.6-0.8)\, {\rm GeV} $, which works well for the hidden-charm pentaquark states.  In this article, we also take the continuum threshold parameters
$\sqrt{s_{0}}=M_{P}+(0.6-0.8)\, {\rm GeV}$ as an additional constraint.

In the present QCD sum rules, we choose the Borel parameters $T^{2}$ and continuum threshold parameters $s_{0}$ to satisfy the
following four criteria:

$\bf{1_\cdot}$ Pole dominance at the phenomenological side;

$\bf{2_\cdot}$ Convergence of the operator product expansion;

$\bf{3_\cdot}$ Appearance of the Borel platforms;

$\bf{4_\cdot}$ Satisfying the energy scale formula.

Now we search for the optimal Borel parameters $T^{2}$ and continuum threshold parameters $s_{0}$ by try and error. The resulting Borel
parameters $T^{2}$, continuum threshold parameters $s_{0}$, pole contributions, and contributions of the vacuum condensates of
dimension 9 and 10 are shown explicitly in Table 1, where the quantum numbers of the hidden-charm pentaquark states are shown explicitly. From Table 1, we can see that the criteria $\bf{1}$ and  $\bf{2}$ of the QCD sum rules are
satisfied.

 We take into account all uncertainties of the input parameters, and obtain the values of the masses and
pole residues of the hidden-charm pentaquark states, which are shown  explicitly in Table 2 and Figs.1-2.

From Table 2 and Fig.1-2, we can see that the  criteria $\bf{3}$ and  $\bf{4}$ of the QCD sum rules are also satisfied. Now the four criteria of the QCD
sum rules are all satisfied, and we expect to make reasonable predictions. The present predictions can be confronted to the experimental data
in the future.

The following two-body strong decays are Okubo-Zweig-Iizuka super-allowed,
\begin{eqnarray}
P_{uusc\bar{c}}\left({\frac{3}{2}}^\pm\right) &\to& \Sigma^+ J/\psi  \, , \,\Xi_c^+ \bar{D}^{0}\, , \, \Sigma_c^{++}{D}_s^- \, , \\
P_{ussc\bar{c}}\left({\frac{3}{2}}^\pm\right) &\to& \Xi^0 J/\psi  \, , \, \,\Omega_c^+\bar{D}^{-}\, ,\, \Xi_c^+{D}_s^-\, .
\end{eqnarray}
We can search for those $P_{c}$ states in the $\Sigma^+J/\psi$, $\Xi_c^+ \bar{D}^{0}$, $\Sigma_c^{++}{D}_s^- $,  $\Xi^0 J/\psi$, $\Omega_c^+ \bar{D}^{-}$,
$\Xi_c^+{D}_s^-$ mass spectrum in the future.

\begin{table}
\begin{center}
\begin{tabular}{|c|c|c|c|c|c|c|c|c|} \hline \hline
                                       &$T^{2}(\rm{GeV}^2)$   &$\sqrt{s_{0}}(\rm{GeV})$   &pole                 &$D_{9}$       &$D_{10}$  \\  \hline
   $P_{uusc\bar{c}}({\frac{3}{2}}^-)$  &$3.4-3.8$             &$5.20\pm0.10$              &$(40-61)\%$          &$(8-11 )\%$   &$(1-2)\%$ \\  \hline
   $P_{ussc\bar{c}}({\frac{3}{2}}^-)$  &$3.6-4.0$             &$5.30\pm0.10$              & $(42-62)\%$         &$( 10-14)\%$  &$\sim1\%$\\  \hline
   $P_{uusc\bar{c}}({\frac{3}{2}}^+)$  &$3.3-3.7$             &$5.30\pm0.10$              &$(40-62)\%$          &$(4-6 )\%$    &$(2-3 )\%$   \\  \hline
   $P_{ussc\bar{c}}({\frac{3}{2}}^+)$  &$3.4-3.8$             &$5.40\pm0.10$              & $(42-63)\%$         &$(5-7 )\%$    &$(1-2 )\%$\\  \hline \hline
\end{tabular}
\end{center}
 \caption{ The Borel parameters, continuum threshold parameters,  pole contributions,  contributions of the vacuum condensates of
dimension 9 and dimension 10 of the hidden-charm  pentaquark states.}
\end{table}

\begin{table}
\begin{center}
\begin{tabular}{|c|c|c|c|c|c|c|c|c|}\hline\hline
                                     &$\mu(\rm{GeV})$  &$M_{P}(\rm{GeV})$     &$\lambda_{P}(\rm{GeV}^6)$  \\  \hline
$P_{uusc\bar{c}}({\frac{3}{2}}^-)$   &2.65             &$4.49\pm0.04$         &$(1.85\pm0.14)\times10^{-3}$ \\ \hline
$P_{ussc\bar{c}}({\frac{3}{2}}^-)$   &2.80             &$4.60\pm0.04$         &$(2.33\pm0.18)\times10^{-3}$  \\  \hline
$P_{uusc\bar{c}}({\frac{3}{2}}^+)$   &2.80             &$4.61\pm0.08$         &$(0.80\pm0.10)\times10^{-3}$  \\  \hline
$P_{ussc\bar{c}}({\frac{3}{2}}^+)$   &3.00             &$4.72\pm0.04$         &$(1.03\pm0.09)\times10^{-3}$  \\  \hline
  \hline
\end{tabular}
\end{center}
\caption{ The energy scales, masses and pole residues of the hidden-charm  pentaquark states. }
\end{table}
\begin{figure}
 \centering
 \includegraphics[totalheight=5cm,width=7cm]{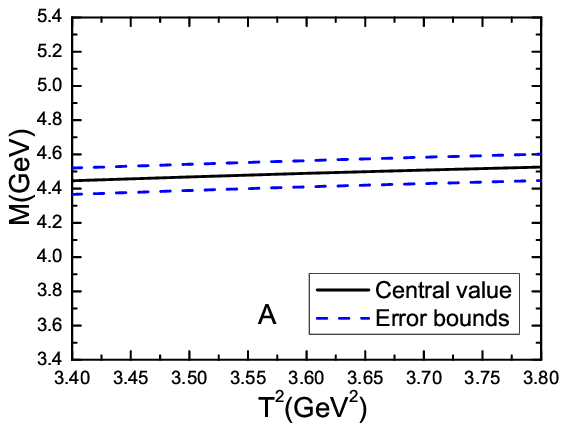}
 \includegraphics[totalheight=5cm,width=7cm]{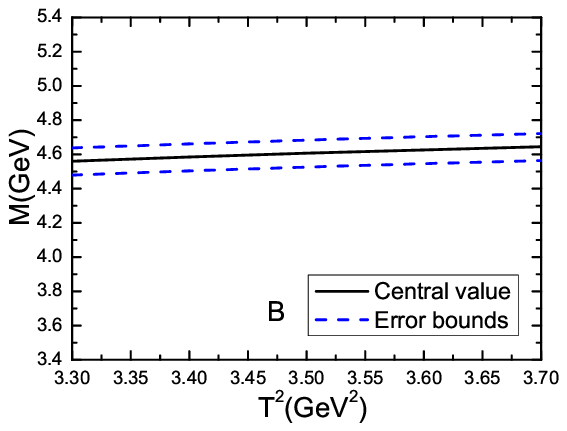}
 \includegraphics[totalheight=5cm,width=7cm]{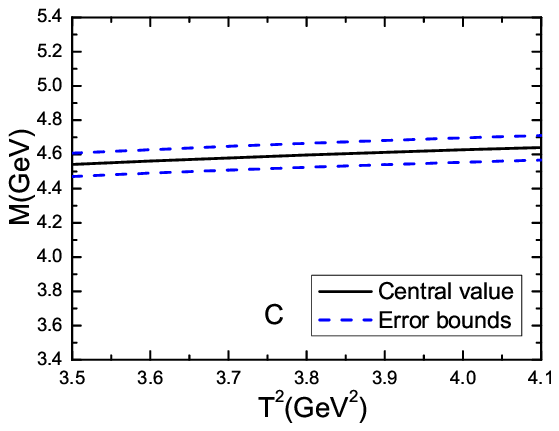}
 \includegraphics[totalheight=5cm,width=7cm]{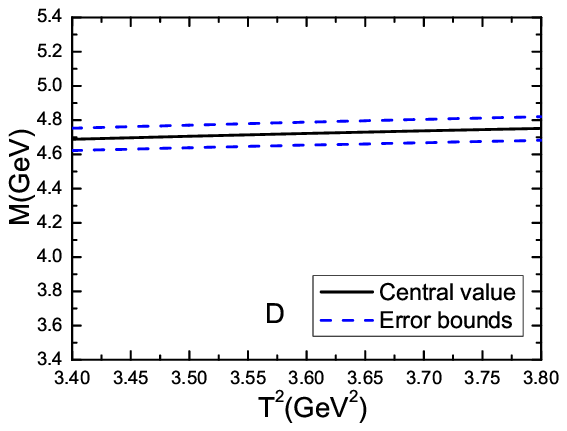}
    \caption{ The masses of the hidden-charm  pentaquark states with
variations  of the Borel parameters $T^{2}$, where the $A$, $B$, $C$ and $D$
denote the pentaquark states $P_{uusc\bar{c}}({\frac{3}{2}}^-)$ , $P_{uusc\bar{c}}({\frac{3}{2}}^+)$, $P_{ussc\bar{c}}({\frac{3}{2}}^-)$ and  $P_{ussc\bar{c}}({\frac{3}{2}}^+)$, respectively.
}
\end{figure}

\begin{figure}
 \centering
 \includegraphics[totalheight=5cm,width=7cm]{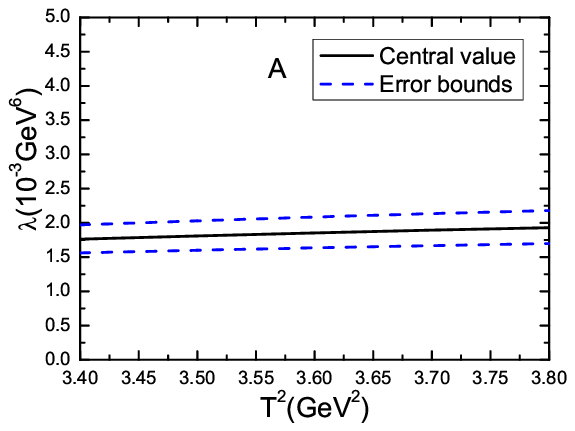}
 \includegraphics[totalheight=5cm,width=7cm]{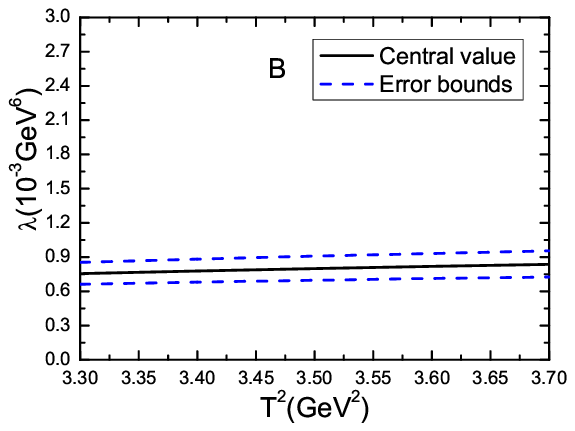}
 \includegraphics[totalheight=5cm,width=7cm]{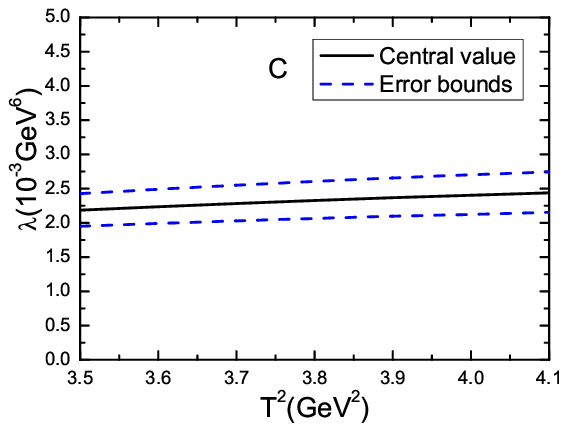}
 \includegraphics[totalheight=5cm,width=7cm]{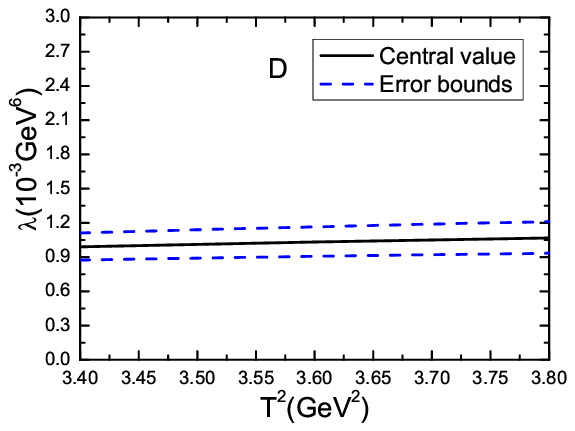}
 \caption{ The pole residues  of the hidden-charm  pentaquark states with
variations of the Borel parameters $T^{2}$, where the $A$, $B$, $C$ and $D$
denote the pentaquark states $P_{uusc\bar{c}}({\frac{3}{2}}^-)$ , $P_{uusc\bar{c}}({\frac{3}{2}}^+)$, $P_{ussc\bar{c}}({\frac{3}{2}}^-)$ and  $P_{ussc\bar{c}}({\frac{3}{2}}^+)$, respectively.
}
\end{figure}

\section{Conclusion}
 In this article, we construct the $S_L-A_H-\bar{c}$ type interpolating currents to study the hidden-charm pentaquark states with $J^{P}=\frac{3}{2}^{\pm}$  and strangeness $S=-1,\,-2$ by calculating the contributions of the vacuum condensates up to dimension-10 in the operator product expansion.
  In calculations, we use the formula $\mu=\sqrt{M_{P}^2-{(2\mathbb{M}_{c}})^2}$ to
determine the ideal energy scales of the QCD spectral densities in a consistent way. We
obtain   the masses and  pole residues of the
hidden-charm pentaquark states with the strangeness $S=-1$, $-2$,   the predicted masses  can be confronted to the experimental data in the
future. On the other hand, we can take the pole residues as basic input parameters to
study relevant processes  of the hidden-charm pentaquark states with the
three-point QCD sum rules.

\section*{Acknowledgements}
This work is supported by   National Natural Science Foundation, Grant Number 11375063; the Fundamental Research Funds for the
Central Universities, Grant Number 2016MS155.

\section*{Appendix}
The QCD spectral densities $\rho^{i1}_j(s)$ and
$\widetilde{\rho}^{i0}_j(s)$ of the hidden-charm pentaquark states with
$j=0,\,3,\,4,\,5,\,6,\,8,\,9,\,10$  are shown as follows,

\begin{eqnarray}
 \notag\ \rho _{0}^{11}(s)&=&\frac{1}{491520{\pi}^{8}}\int dydz \, yz(1-y-z)^4(s-\hat{m}_c^2)^4
 (7s-2\hat{m}_c^2) \, ,\\
 \widetilde{\rho} _{0}^{10}(s)&=&\frac{1}{983040{\pi}^{8}}\int dydz\, (y+z)(1-y-z)^4(s-\hat{m}_c^2)^4
 (6s-\hat{m}_c^2)\, ,\\
 \notag\
 \\ \notag\ \rho _{3}^{11}(s)&=&-\frac{m_{c}\left\langle\bar{q}q\right\rangle}{3072{\pi}^6}\int dydz\, (y+z)(1-y-z)^2(s-\hat{m}_c^2)^3 \\
 \notag\  &&-\frac{2m_{s}\left\langle\bar{q}q\right\rangle-m_{s}\left\langle\bar{s}s\right\rangle}{3072{\pi}^6}\int dydz \, yz(1-y-z)^2(s-\hat{m}_c^2)^2(5s-2\hat{m}_c^2) \, ,\\
  \notag\ \widetilde{\rho} _{3}^{10}(s)&=&-\frac{m_c\left\langle\bar{q}q\right\rangle}{1536{\pi}^6}\int dydz \, (1-y-z)^2(s-\hat{m}_c^2)^3 \\
  &&-\frac{2m_{s}\left\langle\bar{q}q\right\rangle-m_{s}\left\langle\bar{s}s\right\rangle}{6144{\pi}^6}\int dydz \, (y+z)(1-y-z)^2(s-\hat{m}_c^2)^2(4s-\hat{m}_c^2) \, ,\\
    \notag\ \rho _{4}^{11}(s)&=&-\frac{m_c^2}{73728{\pi}^6}\left\langle\frac{{{\alpha}_s}GG}{{\pi}}\right\rangle\int dydz \,\frac{z^3+y^3}{y^2z^2}(1-y-z)^4(s-\hat{m}_c^2)(2s-\hat{m}_c^2) \\
   \notag\ &&-\frac{19}{7077888{\pi}^6}\left\langle\frac{{{\alpha}_s}GG}{{\pi}}\right\rangle\int dydz \,  (y+z)(1-y-z)^3(s-\hat{m}_c^2)^2(7s-4\hat{m}_c^2)\\
\notag\ &&+\frac{13}{393216{\pi}^6}\left\langle\frac{{{\alpha}_s}GG}{{\pi}}\right\rangle\int dydz \, yz(1-y-z)^2(s-\hat{m}_c^2)^2(5s-2\hat{m}_c^2)\\
 \notag\ &&-\frac{3{m_s}m_c}{131072{\pi}^6}\left\langle\frac{{{\alpha}_s}GG}{{\pi}}\right\rangle\int dydz \, (y+z)(1-y-z)(s-{\hat{m}_c^2})^2 \, ,\\
 \notag\ \widetilde{\rho} _{4}^{10}(s)&=&\frac{1}{147456{\pi}^6}\left\langle\frac{{{\alpha}_s}GG}{{\pi}}\right\rangle\int dydz \, \frac{z^3+y^3}{y^2z^2}(1-y-z)^4(s-{\hat{m}}_c^2)(2s^2-4{\hat{m}}_c^2s+{\hat{m}}_c^4)\\
 \notag\
  \end{eqnarray}
 \begin{eqnarray}
  \notag\ &&-\frac{19}{1179648{\pi}^6}\left\langle\frac{{{\alpha}_s}GG}{{\pi}}\right\rangle\int dydz \, (1-y-z)^3(s-{\hat{m}}_c^2)^2(2s-{\hat{m}}_c^2)\\
 \notag\ &&+\frac{13}{786432{\pi}^6}\left\langle\frac{{{\alpha}_s}GG}{{\pi}}\right\rangle\int dydz \,  (y+z)(1-y-z)^2(s-{\hat{m}}_c^2)^2(4s-{\hat{m}}_c^2)\\
 &&-\frac{3{m_s}m_c}{65536{\pi}^6}\left\langle\frac{{{\alpha}_s}GG}{{\pi}}\right\rangle\int dydz\, (1-y-z)(s-{\hat{m}}_c^2)^2\, ,\\
 \notag\ \rho _{5}^{11}(s)&=&\frac{m_c\left\langle\bar{s}g_s{\sigma}Gs\right\rangle}{196608{\pi}^6}\int dydz \, \frac{z^2+y^2}{yz}(1-y-z)^2(1+2y+2z)(s-{\hat{m}}_c^2)^2\\
 \notag\
\notag\ &&+\frac{19m_c\left\langle\bar{q}g_s{\sigma}Gq\right\rangle}{32768{\pi}^6}\int dydz \, (y+z)(1-y-z)(s-{\hat{m}}_c^2)^2\\
\notag\ &&+\frac{57m_s\left\langle\bar{q}g_s{\sigma}Gq\right\rangle-16m_s\left\langle\bar{s}g_s{\sigma}Gs\right\rangle}{24576{\pi}^6}\int dydz \, yz(1-y-z)(s-{\hat{m}}_c^2)(2s-{\hat{m}}_c^2)\, ,\\
\notag\ \widetilde{\rho} _{5}^{10}(s)&=&\frac{m_c\left\langle\bar{s}g_s{\sigma}Gs\right\rangle}{196608{\pi}^6}\int dydz \, \frac{y+z}{yz}(1-y-z)^2(1+2y+2z)(s-{\hat{m}}_c^2)^2\\
\notag\ &&+\frac{19m_c\left\langle\bar{q}g_s{\sigma}Gq\right\rangle}{16384{\pi}^6}\int dydz \, (1-y-z)(s-{\hat{m}}_c^2)^2\\
\notag\ &&+\frac{57m_s\left\langle\bar{q}g_s{\sigma}Gq\right\rangle-16m_s\left\langle\bar{s}g_s{\sigma}Gs\right\rangle}{98304{\pi}^6}\int dydz \, (y+z)(1-y-z)(s-{\hat{m}}_c^2)(3s-{\hat{m}}_c^2)\, ,\\
\\
\notag\ \rho _{6}^{11}(s)&=&\frac{\left\langle\bar{s}s\right\rangle\left\langle\bar{q}q\right\rangle}{96{\pi}^4}\int dydz \, yz(1-y-z)(s-{\hat{m}}_c^2)(2s-{\hat{m}}_c^2)\\
\notag\ &&+\frac{2{m_s}m_c\left\langle\bar{q}q\right\rangle^2-{m_s}m_c\left\langle\bar{s}s\right\rangle\left\langle\bar{q}q\right\rangle}{384{\pi}^4}\int dydz \, (y+z)(s-{\hat{m}}_c^2)\, ,\\
\notag\ \widetilde{\rho} _{6}^{10}(s)&=&\frac{\left\langle\bar{s}s\right\rangle\left\langle\bar{q}q\right\rangle}{384{\pi}^4}\int dydz \, (y+z)(1-y-z)(s-{\hat{m}}_c^2)(3s-{\hat{m}}_c^2)\\
&&+\frac{2{m_s}m_c\left\langle\bar{q}q\right\rangle^2-{m_s}m_c\left\langle\bar{s}s\right\rangle\left\langle\bar{q}q\right\rangle}{192{\pi}^4}\int dydz \, (s-{\hat{m}}_c^2)\, ,\\
\notag\ \rho _{8}^{11}(s)&=&-\frac{16\left\langle\bar{s}g_s{\sigma}Gs\right\rangle\left\langle\bar{q}q\right\rangle+19\left\langle\bar{q}g_s{\sigma}Gq\right\rangle\left\langle\bar{s}s\right\rangle}{6144{\pi}^4}\int dydz \, yz(3s-2{\hat{m}}_c^2)\\
\notag\ &&-\frac{\left\langle\bar{s}g_s{\sigma}Gs\right\rangle\left\langle\bar{q}q\right\rangle}{12288{\pi}^4}\int dydz \, (y+z)(1-y-z)(5s-4{\hat{m}}_c^2)\\
\notag\ &&-\frac{192{m_s}m_c\left\langle\bar{q}g_s{\sigma}Gq\right\rangle\left\langle\bar{q}q\right\rangle-32{m_s}m_c\left\langle\bar{s}g_s{\sigma}Gs\right\rangle\left\langle\bar{q}q\right\rangle-57{m_s}m_c\left\langle\bar{q}g_s{\sigma}Gq\right\rangle\left\langle\bar{s}s\right\rangle}{73728{\pi}^4}\sqrt{1-4m_c^2/s}\, ,\\
\notag\ \widetilde{\rho} _{8}^{10}(s)&=&-\frac{16\left\langle\bar{s}g_s{\sigma}Gs\right\rangle\left\langle\bar{q}q\right\rangle+19\left\langle\bar{q}g_s{\sigma}Gq\right\rangle\left\langle\bar{s}s\right\rangle}{12288{\pi}^4}\int dydz \, (y+z)(2s-{\hat{m}}_c^2)\\
\notag\ &&-\frac{{192m_s}m_c\left\langle\bar{q}g_s{\sigma}Gq\right\rangle\left\langle\bar{q}q\right\rangle-32{m_s}m_c\left\langle\bar{s}g_s{\sigma}Gs\right\rangle\left\langle\bar{q}q\right\rangle-57{m_s}m_c\left\langle\bar{q}g_s{\sigma}Gq\right\rangle\left\langle\bar{s}s\right\rangle}{36864{\pi}^4}\sqrt{1-4m_c^2/s} \\
 &&-\frac{\left\langle\bar{s}g_s{\sigma}Gs\right\rangle\left\langle\bar{q}q\right\rangle}{6144{\pi}^4}\int dydz \, (1-y-z)(4s-3{\hat{m}}_c^2)\, ,\\
 \notag\ \rho _{9}^{11}(s)&=&-\frac{m_c\left\langle\bar{s}s\right\rangle\left\langle\bar{q}q\right\rangle^2}{144{\pi}^2}\sqrt{1-4m_c^2/s}\, ,\\
\widetilde{\rho} _{9}^{10}(s)&=&-\frac{m_c\left\langle\bar{s}s\right\rangle\left\langle\bar{q}q\right\rangle^2}{72{\pi}^2}\sqrt{1-4m_c^2/s}\, ,\\
\notag\
 \end{eqnarray}
 \begin{eqnarray}
 \notag\ \rho _{10}^{11}(s)&=&\frac{19\left\langle\bar{s}g_s{\sigma}Gs\right\rangle\left\langle\bar{q}g_s{\sigma}Gq\right\rangle}{24576{\pi}^4}\int dy \, y(1-y)
 \left\{2+s{\delta}(s-{\widetilde{m}}_c^2)\right\} \\
 \notag\ &&+\frac{17\left\langle\bar{s}g_s{\sigma}Gs\right\rangle\left\langle\bar{q}g_s{\sigma}Gq\right\rangle}{442368{\pi}^4}\int dydz \,  (y+z)\left\{4+s{\delta}(s-{\hat{m}}_c^2)\right\} \\
  \notag\ &&+\frac{48{m_s}m_c\left\langle\bar{q}g_s{\sigma}Gq\right\rangle^2-11{m_s}m_c\left\langle\bar{s}g_s{\sigma}Gs\right\rangle\left\langle\bar{q}g_s{\sigma}Gq\right\rangle}{147456{\pi}^4}\int dy \, \left(3+\frac{s}{T^2}\right){\delta}(s-{\widetilde{m}}_c^2)\, ,\\
 \notag\ \widetilde{\rho} _{10}^{10}(s)&=&\frac{19\left\langle\bar{s}g_s{\sigma}Gs\right\rangle\left\langle\bar{q}g_s{\sigma}Gq\right\rangle}{49152{\pi}^4}\int dy\, \left\{1+s{\delta}(s-{\widetilde{m}}_c^2)\right\}\\
 \notag\ &&+\frac{17\left\langle\bar{s}g_s{\sigma}Gs\right\rangle\left\langle\bar{q}g_s{\sigma}Gq\right\rangle}{221184{\pi}^4}\int dydz\,  \left\{3+s{\delta}(s-{\hat{m}}_c^2)\right\}\\
  \notag\ &&+\frac{48{m_s}m_c\left\langle\bar{q}g_s{\sigma}Gq\right\rangle^2-19{m_s}m_c\left\langle\bar{s}g_s{\sigma}Gs\right\rangle\left\langle\bar{q}g_s{\sigma}Gq\right\rangle}{73728{\pi}^4}\int dy\,  \left(2+\frac{s}{T^2}\right){\delta}(s-{\widetilde{m}}_c^2)\, ,\\
 \end{eqnarray}
 \begin{eqnarray}
 \notag\ \rho _{0}^{21}(s)&=&\frac{1}{491520{\pi}^{8}}\int dydz \, yz(1-y-z)^4(s-\hat{m}_c^2)^4
 (7s-2\hat{m}_c^2) \\
 \notag\ &&+\frac{m_{s}m_{c}}{49152{\pi}^{8}}\int dydz \, (y+z)(1-y-z)^3(s-\hat{m}_c^2)^4,\\
 \notag\ \widetilde{\rho} _{0}^{20}(s)&=&\frac{1}{983040{\pi}^{8}}\int dydz \, (y+z)(1-y-z)^4(s-\hat{m}_c^2)^4
 (6s-\hat{m}_c^2) \\
 &&+\frac{m_{s}m_{c}}{24576{\pi}^{8}}\int dydz \, (1-y-z)^3(s-\hat{m}_c^2)^4,\\
 \notag\
 \\ \notag\ \rho _{3}^{21}(s)&=&-\frac{m_{c}\left\langle\bar{q}q\right\rangle}{3072{\pi}^6}\int dydz \, (y+z)(1-y-z)^2(s-\hat{m}_c^2)^3 \\
  \notag\  &&+\frac{m_{s}\left\langle\bar{s}s\right\rangle-m_{s}\left\langle\bar{q}q\right\rangle}{1536{\pi}^6}\int dydz \, yz(1-y-z)^2(s-\hat{m}_c^2)^2(5s-2\hat{m}_c^2)\, , \\
 \notag\ \widetilde{\rho} _{3}^{20}(s)&=&-\frac{m_c\left\langle\bar{q}q\right\rangle}{1536{\pi}^6}\int dydz \,  (1-y-z)^2(s-\hat{m}_c^2)^3 \\
 &&+\frac{m_{s}\left\langle\bar{s}s\right\rangle-m_{s}\left\langle\bar{q}q\right\rangle}{3072{\pi}^6}\int dydz \, (y+z)(1-y-z)^2(s-\hat{m}_c^2)^2(4s-\hat{m}_c^2)\, , \\
 \notag\ \rho _{4}^{21}(s)&=&-\frac{m_c^2}{73728{\pi}^6}\left\langle\frac{{{\alpha}_s}GG}{{\pi}}\right\rangle\int dydz \, \frac{y^3+z^3}{y^2z^2}(1-y-z)^4(s-\hat{m}_c^2)(2s-\hat{m}_c^2) \\
\notag\ &&-\frac{m_{s}m_{c}}{73728{\pi}^6}\left\langle\frac{{{\alpha}_s}GG}{{\pi}}\right\rangle\int dydz \, \frac{y^2+z^2}{yz}(1-y-z)^3(s-\hat{m}_c^2)^2(5s-3\hat{m}_c^2)\\
\notag\
&&+\frac{m_{s}m_{c}}{36864{\pi}^6}\left\langle\frac{{{\alpha}_s}GG}{{\pi}}\right\rangle\int
dydz \, \frac{y^3+z^3}{y^2z^2}(1-y-z)^3(s-\hat{m}_c^2)^2\\
\notag\ &&-\frac{m_{s}m^3_{c}}{36864{\pi}^6}\left\langle\frac{{{\alpha}_s}GG}{{\pi}}\right\rangle\int dydz \, \frac{y^2+z^2}{y^2z^2}(1-y-z)^3(s-\hat{m}_c^2)\\
 \notag\ &&-\frac{19}{7077888{\pi}^6}\left\langle\frac{{{\alpha}_s}GG}{{\pi}}\right\rangle\int dydz \, (y+z)(1-y-z)^3(s-\hat{m}_c^2)^2(7s-4\hat{m}_c^2)\\
 \notag\ &&-\frac{5}{98304{\pi}^6}\left\langle\frac{{{\alpha}_s}GG}{{\pi}}\right\rangle\int dydz \, yz(1-y-z)^2(s-\hat{m}_c^2)^2(5s-2\hat{m}_c^2)\\
  \notag\ &&+\frac{{m_s}m_c}{8192{\pi}^6}\left\langle\frac{{{\alpha}_s}GG}{{\pi}}\right\rangle\int dydz \, (y+z)(1-y-z)(s-{\hat{m}_c^2})^2 \\
  \notag\
&&-\frac{m_{s}m_{c}}{131072{\pi}^6}\left\langle\frac{{{\alpha}_s}GG}{{\pi}}\right\rangle\int
dydz \, \frac{y^2+z^2}{yz}(1-y-z)^2(s-\hat{m}_c^2)^2\\
\notag\
&&+\frac{m_{s}m_{c}}{98304{\pi}^6}\left\langle\frac{{{\alpha}_s}GG}{{\pi}}\right\rangle\int
dydz \, \frac{y^2+z^2}{yz}(1-y-z)^3(s-\hat{m}_c^2)(5s-3\hat{m}_c^2)\, ,\\
 \notag\ \widetilde{\rho} _{4}^{20}(s)&=&\frac{1}{147456{\pi}^6}\left\langle\frac{{{\alpha}_s}GG}{{\pi}}\right\rangle\int dydz \, \frac{y^3+z^3}{y^2z^2}(1-y-z)^4(s-{\hat{m}}_c^2)(2s^2-4{\hat{m}}_c^2+{\hat{m}}_c^4)\\
 \notag\
&&+\frac{5m_{s}m_{c}}{73728{\pi}^6}\left\langle\frac{{{\alpha}_s}GG}{{\pi}}\right\rangle\int
dydz \, \frac{y^2+z^2}{y^2z^2}(1-y-z)^3(s-{\hat{m}}_c^2)^2\\
\notag\
&&-\frac{m_{s}m_{c}}{36864{\pi}^6}\left\langle\frac{{{\alpha}_s}GG}{{\pi}}\right\rangle\int
dydz \, \frac{y+z}{yz}(1-y-z)^3(s-{\hat{m}}_c^2)^2\\
\notag\
&&-\frac{m_{s}m_{c}}{36864{\pi}^6}\left\langle\frac{{{\alpha}_s}GG}{{\pi}}\right\rangle\int
dydz \, (1-y-z)^3(2-y-z)(s-{\hat{m}}_c^2)s\\
\notag\ &&-\frac{m_{s}m^3_{c}}{36864{\pi}^6}\left\langle\frac{{{\alpha}_s}GG}{{\pi}}\right\rangle\int dydz \, \frac{y^3+z^3}{y^3z^3}(1-y-z)^3(s-\hat{m}_c^2)\\
 \notag\
 \end{eqnarray}

\begin{eqnarray}
\notag\ &&-\frac{19}{1179648{\pi}^6}\left\langle\frac{{{\alpha}_s}GG}{{\pi}}\right\rangle\int dydz \, (1-y-z)^3(s-{\hat{m}}_c^2)^2(2s-{\hat{m}}_c^2)\\
 \notag\ &&+\frac{1}{73728{\pi}^6}\left\langle\frac{{{\alpha}_s}GG}{{\pi}}\right\rangle\int dydz \, (y+z)(1-y-z)^2(s-{\hat{m}}_c^2)^2(4s-{\hat{m}}_c^2)\\
\notag\
&&+\frac{{m_s}m_c}{4096{\pi}^6}\left\langle\frac{{{\alpha}_s}GG}{{\pi}}\right\rangle\int
dydz \, (1-y-z)(s-{\hat{m}}_c^2)^2 \\
\notag\ &&-\frac{3}{393216{\pi}^6}\left\langle\frac{{{\alpha}_s}GG}{{\pi}}\right\rangle\int dydz \, yz(1-y-z)^2(s-\hat{m}_c^2)^2(4s-\hat{m}_c^2)\\
\notag\
&&-\frac{m_{s}m_{c}}{131072{\pi}^6}\left\langle\frac{{{\alpha}_s}GG}{{\pi}}\right\rangle\int
dydz \, \frac{y+z}{yz}(1-y-z)^2(s-\hat{m}_c^2)^2\\
&&+\frac{m_{s}m_{c}}{49152{\pi}^6}\left\langle\frac{{{\alpha}_s}GG}{{\pi}}\right\rangle\int
dydz \, \frac{y+z}{yz}(1-y-z)^3(s-\hat{m}_c^2)(2s-\hat{m}_c^2)\,  ,\\
\notag\ \rho _{5}^{21}(s)&=&\frac{19m_c\left\langle\bar{s}g_s{\sigma}Gs\right\rangle}{32768{\pi}^6}\int dydz \, (y+z)(1-y-z)(s-{\hat{m}}_c^2)^2\\
\notag\ &&+\frac{3m_s\left\langle\bar{q}g_s{\sigma}Gq\right\rangle-2m_s\left\langle\bar{s}g_s{\sigma}Gs\right\rangle}{1536{\pi}^6}\int dydz \, yz(1-y-z)(s-{\hat{m}}_c^2)(2s-{\hat{m}}_c^2)\\
 \notag\ &&+\frac{m_s\left\langle\bar{q}g_s{\sigma}Gq\right\rangle}{32768{\pi}^6}\int dydz \, (y+z)(1-y-z)^2(s-{\hat{m}}_c^2)(3s-{\hat{m}}_c^2)\\
 \notag\ &&+\frac{m_c\left\langle\bar{q}g_s{\sigma}Gq\right\rangle}{196608{\pi}^6}\int dydz \, \frac{y^2+z^2}{yz}(1-y-z)^2(1+2y+2z)(s-{\hat{m}}_c^2)^2 \, , \\
\notag\ \widetilde{\rho} _{5}^{20}(s)&=&\frac{19m_c\left\langle\bar{s}g_s{\sigma}Gs\right\rangle}{32768{\pi}^6}\int dydz \, (1-y-z)(s-{\hat{m}}_c^2)^2\\
\notag\ &&+\frac{3m_s\left\langle\bar{q}g_s{\sigma}Gq\right\rangle-2m_s\left\langle\bar{s}g_s{\sigma}Gs\right\rangle}{6144{\pi}^6}\int dydz \, (y+z)(1-y-z)(s-{\hat{m}}_c^2)(3s-{\hat{m}}_c^2)\\
\notag\ &&+\frac{m_s\left\langle\bar{q}g_s{\sigma}Gq\right\rangle}{16384{\pi}^6}\int dydz \, (1-y-z)^2(s-{\hat{m}}_c^2)(5s-3{\hat{m}}_c^2)\\
&&-\frac{m_c\left\langle\bar{q}g_s{\sigma}Gq\right\rangle}{196608{\pi}^6}\int dydz \, \frac{y+z}{yz}(1-y-z)^2(1-4y-4z)(s-{\hat{m}}_c^2)^2\, ,\\
\notag\ \rho _{6}^{21}(s)&=&\frac{\left\langle\bar{s}s\right\rangle\left\langle\bar{q}q\right\rangle}{96{\pi}^4}\int dydz \, yz(1-y-z)(s-{\hat{m}}_c^2)(2s-{\hat{m}}_c^2)\\
 \notag\ &&+\frac{4{m_s}m_c\left\langle\bar{s}s\right\rangle\left\langle\bar{q}q\right\rangle-{m_s}m_c\left\langle\bar{s}s\right\rangle^2}{384{\pi}^4}\int dydz \, (y+z)(s-{\hat{m}}_c^2)\, ,\\
  \notag\ \widetilde{\rho} _{6}^{20}(s)&=&\frac{\left\langle\bar{s}s\right\rangle\left\langle\bar{q}q\right\rangle}{384{\pi}^4}\int dydz \, (y+z)(1-y-z)(s-{\hat{m}}_c^2)(3s-{\hat{m}}_c^2)\\
 \notag\ &&+\frac{{4{m_s}m_c\left\langle\bar{s}s\right\rangle\left\langle\bar{q}q\right\rangle-m_s}m_c\left\langle\bar{s}s\right\rangle^2}{192{\pi}^4}\int dydz \, (s-{\hat{m}}_c^2)  \, ,\\
 \notag\ \rho _{8}^{21}(s)&=&-\frac{19\left\langle\bar{s}g_s{\sigma}Gs\right\rangle\left\langle\bar{q}q\right\rangle+16\left\langle\bar{q}g_s{\sigma}Gq\right\rangle\left\langle\bar{s}s\right\rangle}{6144{\pi}^4}\int dydz \, yz(3s-2{\hat{m}}_c^2)\\
    \notag\ &&+\frac{5{m_s}m_c\left\langle\bar{s}g_s{\sigma}Gs\right\rangle\left\langle\bar{s}s\right\rangle-12{m_s}m_c\left\langle\bar{s}g_s{\sigma}Gs\right\rangle\left\langle\bar{q}q\right\rangle-12{m_s}m_c\left\langle\bar{q}g_s{\sigma}Gq\right\rangle\left\langle\bar{s}s\right\rangle}{4608{\pi}^4}\sqrt{1-4m_c^2/s}\\
    \notag\ &&-\frac{\left\langle\bar{q}g_s{\sigma}Gq\right\rangle\left\langle\bar{s}s\right\rangle}{12288{\pi}^4}\int dydz \, (y+z)(1-y-z)(5s-4{\hat{m}}_c^2)\\
   \notag\ &&-\frac{{m_s}m_c\left\langle\bar{q}g_s{\sigma}Gq\right\rangle\left\langle\bar{s}s\right\rangle}{12288{\pi}^4}\int dydz \, \frac{y^2+z^2}{yz}(1-2y-2z)\, , \\
   \notag\ \widetilde{\rho}_{8}^{20}(s)&=&-\frac{19\left\langle\bar{s}g_s{\sigma}Gs\right\rangle\left\langle\bar{q}q\right\rangle+16\left\langle\bar{q}g_s{\sigma}Gq\right\rangle\left\langle\bar{s}s\right\rangle}{12288{\pi}^4}\int dydz \, (y+z)(2s-{\hat{m}}_c^2)\\
  \notag\
\end{eqnarray}

\begin{eqnarray}
    \notag\ &&+\frac{5{m_s}m_c\left\langle\bar{s}g_s{\sigma}Gs\right\rangle\left\langle\bar{s}s\right\rangle-12{m_s}m_c\left\langle\bar{s}g_s{\sigma}Gs\right\rangle\left\langle\bar{q}q\right\rangle-12{m_s}m_c\left\langle\bar{q}g_s{\sigma}Gq\right\rangle\left\langle\bar{s}s\right\rangle}{2304{\pi}^4}\sqrt{1-4m_c^2/s}\\
    \notag\ &&-\frac{\left\langle\bar{q}g_s{\sigma}Gq\right\rangle\left\langle\bar{s}s\right\rangle}{6144{\pi}^4}\int dydz \, (1-y-z)(4s-3{\hat{m}}_c^2)\\
  &&-\frac{{m_s}m_c\left\langle\bar{q}g_s{\sigma}Gq\right\rangle\left\langle\bar{s}s\right\rangle}{12288{\pi}^4}\int dydz \, \frac{y+z}{yz}(1-2y-2z)\, ,\\
   \notag\ \rho _{9}^{21}(s)&=&-\frac{m_c\left\langle\bar{s}s\right\rangle^2\left\langle\bar{q}q\right\rangle}{144{\pi}^2}\sqrt{1-4m_c^2/s}\\
\notag\ &&+\frac{m_s\left\langle\bar{s}s\right\rangle^2\left\langle\bar{q}q\right\rangle}{144{\pi}^2}\int dy \, y(1-y)\{2+s\delta(s-{\widetilde{m}}_c^2)\}\, ,\\
 \notag\ \widetilde{\rho} _{9}^{20}(s)&=&-\frac{m_c\left\langle\bar{s}s\right\rangle^2\left\langle\bar{q}q\right\rangle}{72{\pi}^2}\sqrt{1-4m_c^2/s}\\
 &&+\frac{m_s\left\langle\bar{s}s\right\rangle^2\left\langle\bar{q}q\right\rangle}{288{\pi}^2}\int dy \, \{1+s\delta(s-{\widetilde{m}}_c^2)\}\, ,\\
   \notag\ \\
  \notag\ \rho _{10}^{21}(s)&=&\frac{19\left\langle\bar{s}g_s{\sigma}Gs\right\rangle\left\langle\bar{q}g_s{\sigma}Gq\right\rangle}{24576{\pi}^4}\int dy\, y(1-y)
 \{2+s{\delta}(s-{\widetilde{m}}_c^2)\} \\
  \notag\
&&+\frac{6{m_s}m_c\left\langle\bar{s}g_s{\sigma}Gs\right\rangle\left\langle\bar{q}g_s{\sigma}Gq\right\rangle-m_{s}m_c\left\langle\bar{s}g_s{\sigma}Gs\right\rangle^2}{9216{\pi}^4}\int
dy \, \left(2+\frac{s}{T^2}\right){\delta}(s-{\widetilde{m}}_c^2) \\
\notag\
&&+\frac{17\left\langle\bar{s}g_s{\sigma}Gs\right\rangle\left\langle\bar{q}g_s{\sigma}Gq\right\rangle}{442368{\pi}^4}\int
dydz\,(y+z)
 \left\{4+s\delta(s-{\hat{m}}_c^2)\right\}\\
\notag\
&&-\frac{{m_s}m_c\left\langle\bar{s}g_s{\sigma}Gs\right\rangle\left\langle\bar{q}g_s{\sigma}Gq\right\rangle}{36864{\pi}^4}\int
dy\,\frac{1-y}{y}{\delta}(s-{\widetilde{m}}_c^2) \\
\notag\
&&+\frac{{m_s}m_c\left\langle\bar{s}g_s{\sigma}Gs\right\rangle\left\langle\bar{q}g_s{\sigma}Gq\right\rangle}{36864{\pi}^4}\int
dydz\, \frac{y^2+z^2}{yz}{\delta}(s-{\hat{m}}_c^2) \, ,\\
 \notag\ \widetilde{\rho} _{10}^{20}(s)&=&\frac{19\left\langle\bar{s}g_s{\sigma}Gs\right\rangle\left\langle\bar{q}g_s{\sigma}Gq\right\rangle}{49152{\pi}^4}\int dy\,
y(1-y) \{1+s{\delta}(s-{\widetilde{m}}_c^2)\} \\
\notag\
&&+\frac{6{m_s}m_c\left\langle\bar{s}g_s{\sigma}Gs\right\rangle\left\langle\bar{q}g_s{\sigma}Gq\right\rangle-m_{s}m_c\left\langle\bar{s}g_s{\sigma}Gs\right\rangle^2}{4608{\pi}^4}\int
dy \, \left(1+\frac{s}{T^2}\right){\delta}(s-{\widetilde{m}}_c^2) \\
\notag\
&&+\frac{17\left\langle\bar{s}g_s{\sigma}Gs\right\rangle\left\langle\bar{q}g_s{\sigma}Gq\right\rangle}{221184{\pi}^4}\int
dydz\,
 \left\{3+s\delta(s-{\hat{m}}_c^2)\right\}\\
 \notag\
&&-\frac{{m_s}m_c\left\langle\bar{s}g_s{\sigma}Gs\right\rangle\left\langle\bar{q}g_s{\sigma}Gq\right\rangle}{36864{\pi}^4}\int
dy\, \frac{1}{y}{\delta}(s-{\widetilde{m}}_c^2)\\
&&+\frac{{m_s}m_c\left\langle\bar{s}g_s{\sigma}Gs\right\rangle\left\langle\bar{q}g_s{\sigma}Gq\right\rangle}{36864{\pi}^4}\int
dydz\, \frac{y+z}{yz}{\delta}(s-{\hat{m}}_c^2)\, ,
\end{eqnarray}
$\int dydz=\int_{y_i}^{y_f}dy\int_{z_i}^{1-y}dz$, $y_f=\frac{1+\sqrt{1-4m_c^2/s}}{2}$, $y_i=\frac{1-\sqrt{1-4m_c^2/s}}{2}$, $z_i=\frac{ym_c^2}{ys-m_c^2}$,  ${\hat{m}_c^2}=\frac{(y+z)m_c^2}{yz}$, ${\widetilde{m}}_c^2=\frac{m_c^2}{y(1-y)}$,  $\int_{y_i}^{y_f}dy \rightarrow \int_0^1 dy$, $\int_{z_i}^{1-y}dz\rightarrow\int_0^{1-y}dz$, when the $\delta$ functions $\delta(s-{\hat{m}_c^2})$ and $\delta(s-{\widetilde{m}}_c^2)$ appear.\\

\end{document}